\documentclass[floatfix,aps,twocolumn,10pt,superscriptaddress,prapplied]{revtex4-2}

\usepackage{bm}
\usepackage{graphicx}
\usepackage{subfigure}
\usepackage{wrapfig}
\usepackage{epstopdf}
\usepackage{siunitx}
\usepackage{braket}
\usepackage{tabularx}
\usepackage{blindtext}
\usepackage{amsmath}
\usepackage{float}

\usepackage{color}
\usepackage[section]{placeins}

\def\bem#1{\begin{mathletters}\label{#1}}
\def\eml{\end{mathletters}}

\def\4#1{{\boldsymbol{#1}}}
\def\8#1{{\widetilde{#1}}}

\begin{document}

\title {Exploring methods for creation of Boron-vacancies in hexagonal Boron Nitride exfoliated from bulk crystal}

\author{T. Zabelotsky}
\affiliation{The Center for Nanoscience and Nanotechnology, The Hebrew University of Jerusalem, Jerusalem 91904, Israel}
\affiliation{Dept. of Applied Physics, Rachel and Selim School of Engineering, Hebrew University, Jerusalem 91904, Israel}

\author{S. Singh} 
\affiliation{The Racah Institute of Physics, The Hebrew University of Jerusalem, Jerusalem 91904, Israel}

\author{G. Haim} 
\affiliation{Dept. of Applied Physics, Rachel and Selim School of Engineering, Hebrew University, Jerusalem 91904, Israel}
\affiliation{School of Physics, The University of Melbourne, Parkville, Victoria 3010, Australia}

\author{R. Malkinson}
\affiliation{The Center for Nanoscience and Nanotechnology, The Hebrew University of Jerusalem, Jerusalem 91904, Israel}
\affiliation{Dept. of Applied Physics, Rachel and Selim School of Engineering, Hebrew University, Jerusalem 91904, Israel}

\author{S. Kadkhodazadeh} 
\affiliation{DTU Nanolab, Technical University of Denmark, Fysikvej, Kongens Lyngby 2800, Denmark }

\author{I. P. Radko}
\affiliation{Department of Physics, Technical University of Denmark, Kongens Lyngby 2800, Denmark}

\author{I. Aharonovich}
\affiliation{School of Mathematical and Physical Sciences, University of Technology Sydney, Ultimo, New South Wales 2007, Australia}
\affiliation{ARC Centre of Excellence for Transformative Meta-Optical Systems (TMOS), Faculty of Science, University of Technology Sydney, Australia}

\author{H. Steinberg}
\affiliation{The Center for Nanoscience and Nanotechnology, The Hebrew University of Jerusalem, Jerusalem 91904, Israel}
\affiliation{The Racah Institute of Physics, The Hebrew University of Jerusalem, Jerusalem 91904, Israel}

\author{Kirstine Berg-S{\o}rensen}
\affiliation{Department of Health Technology, Technical University of Denmark, Kongens Lyngby 2800 Denmark}

\author{A. Huck}
\affiliation{Center for Macroscopic Quantum States (bigQ), Department of Physics, Technical University of Denmark, 2800 Kongens Lyngby,
Denmark}

\author{T. Taniguchi}
\affiliation{International Center for Materials Nanoarchitectonics, National Institute for Materials Science, 1-1 Namiki,
Tsukuba 305-0044, Japan}

\author{K. Watanabe}
\affiliation{Research Center for Functional Materials, National Institute for Materials Science, 1-1 Namiki, Tsukuba 305-0044, Japan}

\author{N. Bar-Gill}

\affiliation{The Center for Nanoscience and Nanotechnology, The Hebrew University of Jerusalem, Jerusalem 91904, Israel}
\affiliation{Dept. of Applied Physics, Rachel and Selim School of Engineering, Hebrew University, Jerusalem 91904, Israel}
\affiliation{The Racah Institute of Physics, The Hebrew University of Jerusalem, Jerusalem 91904, Israel}

\begin{abstract}
  Boron vacancies (VB${^-}$) in hexagonal boron-nitride (hBN) have sparked great interest in recent years, due to their electronic spin properties. 
  Since hBN can be readily integrated into devices where it interfaces a huge variety of other 2D materials, boron vacancies may serve as a precise sensor which can be deployed at very close proximity to many important materials systems. 
  Boron vacancy defects may be produced by a number of existing methods, the use of which may depend on the final application. Any method should reproducibly generate defects with controlled  density and desired pattern. 
  To date, however, detailed studies of such methods are missing. 
  In this paper we study various techniques, focused ion beam (FIB), electron irradiation and ion implantation, for the preparation of hBN flakes from bulk crystals, and relevant post-processing treatments to create VB${^-}$s as a function of flake thickness and defect concentrations. 
  We find that flake thickness plays an important role when optimising implantation parameters, while careful sample cleaning proved important to achieve best results.
\end{abstract}

\maketitle

\section{Introduction}
Hexagonal boron-nitride (hBN) is a graphite-analog layered material with a hexagonal lattice where Boron and Nitrogen atoms alternate in each hexagon. hBN is a wide band-gap insulator, which became an essencial ingredient in stacked van der Waals (vdW) heterostructures\cite{Geim2013}.
hBN was first used in vdW heterostructures as an ultra-flat, electrically stable substrate \cite{Dean2010}, and as a means to fully encapsulate graphene \cite{Mayorov_2011}. It was soon realized that ultrathin hBN flakes can be used as tunnel barriers \cite{Amet_2012_PhysRevB.85.073405}, and that hBN capping can stabilize air-sensitive materials \cite{chao_2015_doi:10.1021/acs.nanolett.5b00648}.
Since hBN has strong adhesion to graphene, it has enabled the use of selective tearing and stacking, the key method used in the assembly of twist-angle-controlled devices \cite{Cao2018}.

Defects in hBN have been studied using electronic transport - where they are identified as nanoscale quantum dots, sensitive to the spectra and compressibility of nearby samples \cite{Keren2020,Greenaway2018}.
They have also been extensively studied using optical fluorescence. 
Color centers in hBN, specifically VB$^-$s, are stable spin defects in the hBN lattice with similar traits to the well known Nitrogen Vacancies (NV$^-$) in diamond \cite{nvSignalSensingReview, Taylor2008, Dolde2011, Pham_2011} and have comparable applications \cite{hbn_vs_other_solids}.

Given the potential of such optically-active spin defects in vdW materials, significant effort has been invested over the past few years in the controlled fabrication and characterization of these color centers \cite{hbnFIB,eIrradiationMeV,vbIonImplant}. Nevertheless, a systematic study of VB$^-$ creation as a function of method (electron irradiation vs. ion implantation), preparation protocol, and flake thickness is still missing.

In this work we have tested different methods for generating VB$^-$s in bulk exfoliated hBN, while focusing on flake preparation and the effect of flake thickness on the yield of VB$^-$s. We describe in detail the preparation method applied to the flakes, the methods applied for VB creation, the characterization of the flakes by fluorescence spectroscopy and spin resonance measurements. We present successful creation of VB$^-$s using 12 keV ion implantation through FIB, with both Nitrogen and Oxygen ion beams. The defects were identified by their spectral signature with a peak emission at $\sim800$ nm and their spin resonance at $~3.47$ GHz at zero field \cite{vbOdmr, Ivady2020}. We observe increasing fluorescence with increasing ion fluence, and identify a thickness dependence, such that thicker flakes show higher fluorescence intensities for the same fluence. Stopping Range of Ions in Matter (SRIM) simulations were performed to better understand the defect creation process. Our results indicate that in practice, vacancies are created in the sample further than the ion stopping range, perhaps due to ion channeling, which is not taken into account during the simulations. This extends the interaction depth of the ion beam with the samples, resulting in a strong thickness dependent fluorescence of the flakes, as observed. In contrast, our attempts at generating VB$^-$ using electron beam in transmission electron microscope (TEM) or electron lithography were not effective.

\section{Preparing hBN flakes}
One of the simplest techniques for preparing hBN flakes is exfoliation from bulk crystal (using blue tape, Nitto), similar to the processes originally developed for graphene and molybdenum disulphide (MoS$_2$) \cite{Geim2013}. It must be noted, that for our purposes the exfoliation should be handled carefully in a clean environment, as it could introduce unwanted carbon contaminants to the flakes \cite{Huang2020}, which could then impede the creation of VB$^-$s, introduce other defects to the material, and lead to decreased Signal to Noise Ratio (SNR) for VB$^-$ spin readout.

Once exfoliated using the tape, flakes were transferred to a Silicon (Si) substrate with a thin layer (285 nm) of Silicon dioxide (SiO2) and gold markers. To clean the flakes from contaminants that were introduced during exfoliation, the samples were heated in forming gas (95\% Ar and 5\% H) for 4 hours at 375$^{\circ}$C. We performed this cleaning process on flakes marked N1, N2, O1 and O2 (see Table~\ref{table:FIB_treatment} below). Additionally, another sample, flake N3, was cleaned using an open air hot plate, for 15 minutes at 400$^{\circ}$C, the results of the characterization for this flake are detailed in the supplementary material \cite{suppRef}.

\section{Defect creation}
\subsection{FIB treatment}
Ion implantation \cite{hbnFIB}, was the focus of the work presented here, and was implemented on flakes N1, N2, N3, O1 and O2, summarized in Table \ref{table:FIB_treatment}. This approach, which resulted in our most successful VB$^-$ yield, was based on using a focused ion beam (FIB, Helios 5 Hydra Dualbeam plasma FIB/SEM instrument), utilizing Nitrogen (N) and Oxygen (O) ion elements at 12 keV.

\begin{table}[tbh]
  \begin{tabular} {c|c|c|c|c} 
      Flake & Flake     & FIB                  & Cleaning & Color \\
      name  & thickness & fluence              & method   & in figures \\
            & [nm]      &$[{\frac{ion}{cm^2}]}$&          &      \\
      \hline
      N1 & 100 / 150 & $6.25\cdot10^{14}$ & forming gas & Blue / Orange \\
      N2 & 73        & dose test          & forming gas & Yellow \\
      N3 & N.A.      & dose test          & hot plate   & N.A. \\
      O1 & 114       & $1.8\cdot10^{13}$  & forming gas & Red \\
      O2 & 83        & dose test          & forming gas & Green \\
      \hline
  \end{tabular}
\caption{FIB testament description for different flakes discussed in this manuscript.}
\label{table:FIB_treatment}
\end{table}

Dose tests were performed with both Oxygen and Nitrogen ion beams. In both cases the parameters used, ion beam energy, beam size, dwell time and beam current, were kept constant, while changing the number of exposure repetitions to achieve varying dosages. The VB$^-$ yield is not sensitive to changing these parameters, as long as the beam fluence is kept unchanged.

The fluence was calculated in the following way:
\begin{equation}%
\large 
Dose = \frac{I_{beam}t_{dwell}N_{repeats}}{qA_{beam}},
\end{equation}
where $I_{beam}$ is the beam current, $t_{dwell}$ is the dwell time, $N_{repeats}$ is the number of repetitions, $q$ is an electron's charge and $A_{beam}$ is the beam size area.

Flakes N1 and O1 underwent a uniform FIB exposure, with fluences of $6.25{\cdot}10^{14}$ $cm^{-2}$ and $1.81{\cdot}10^{13}$ $cm^{-2}$ respectively, while the flakes N2, N3 and O2 underwent dose tests in an array of square regions of size $3{\times}3 \ {\mu}m^2$.

We note that increasing the fluence further (see supplementary material \cite{suppRef}), to the range of $\sim 10^{16}$ $cm^{-2}$, creates visible damage on the flakes (observed optically as a change in the flake's color), although VB$^-$s were created. Increasing the dose even further, up to $10^{17}-10^{18}$ $cm^{-2}$ could sputter away layers without creating any VB$^-$s leading to unwanted damage to the flakes, although we note that there are other factors that should be taken into account in this context such as flake thickness and beam energy.

\subsection{Other methods}
In addition to FIB, we attempted ion implantation using standard, commercial (semiconductor industry) implanter instrumentation (Innovion Corp.) to reproduce previously reported results \cite{vbIonImplant}. In our case these trials were not successful, resulting in measured fluorescence but not with characteristics matching those of VB$^-$s. We attribute the measured fluorescence to other defects. Nevertheless, we note that for this process no cleaning was performed on the flakes post transfer to the Si substrate (before implantation). Since unclean flakes exhibited limited defect creation also using FIB, we deduce that sample cleaning before implantation plays an important role.

Furthermore, we studied electron irradiation instead of ion implantation, performed using two electron sources: e-beam lithography (100keV, Elionix ELS-G100) and TEM (80keV and 300keV, Themis-Z).
The hBN samples used for these electron irradiation studies went through the same sample preparation as detailed for the FIB. For TEM irradiation, the flakes were transferred to TEM Gold grids (200 Mesh) with carbon film (R 1.2/1.3) using the following procedure:
first, hBN flakes were transferred to a Si substrate spin coated with Poly(vinyl alcohol) (PVA), then the TEM grid's carbon side was placed on the Si, and finally a drop of water was added to dissolve the PVA, allowing the hBN flakes to be transferred to the grid.

Electron irradiation under the above conditions did not result in successful creation of VB$^-$s. We attribute this to insufficient energy of the electron beam. Theoretically, the maximum energy that 80 keV and 100 keV electrons can transfer to B atoms are approximately 17.5 eV and 22.2 eV, respectively~\cite{Egerton3rded,Egerton2004}. Given the displacement threshold of 19.36 eV reported for B in pristine hBN~\cite{Kotakoski2010}, it is unlikely to create VB$^-$s in hBN flakes using 80 keV electrons. Although the energy transfer from 100 keV electrons should be sufficient for VB$^-$s creation, the probability of such a process is relatively low and likely requires very high electron beam doses~\cite{Kotakoski2010}.   

\section{Sample characterization}

\subsection{Optical spectroscopy}

\begin{figure}[bth!]
{\includegraphics[width = 0.5 \linewidth]{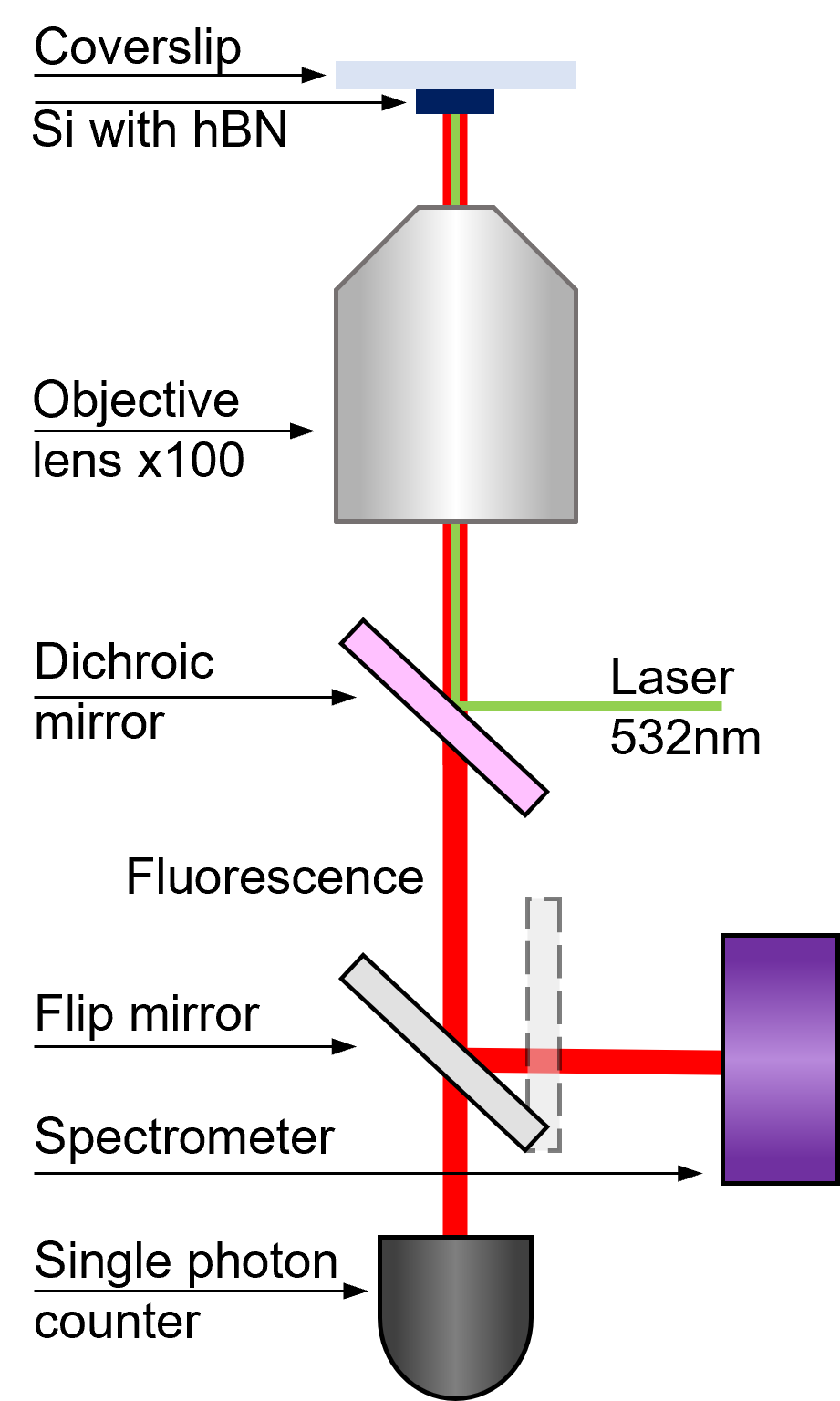}}
\caption{Confocal setup sketch for spectral measurements, 532 nm laser is used for excitation through an Olympus air objective M Plan Apo 100x/0.95. PL is collected through a dichroic mirror either to a single photon counter (Perkin-Elmer SPCM, model AQRH-14) for confocal scanning, or to a spectrometer (Andor SR500i).}
\label{fig:Spectral_setup}
\end{figure}

Given the well known spectral signature of VB$^-$ defects \cite{vbOdmr}, optical spectroscopy is an efficient and convenient tool for characterizing defect creation.

Emission spectra of the defects was measured using a home made confocal setup (Figure \ref{fig:Spectral_setup}) equipped with a spectrometer (Andor SR500i). 

\begin{figure}[bht!]
\centering
    \includegraphics[width = 1 \linewidth, trim={0 0.25cm 0 0}]{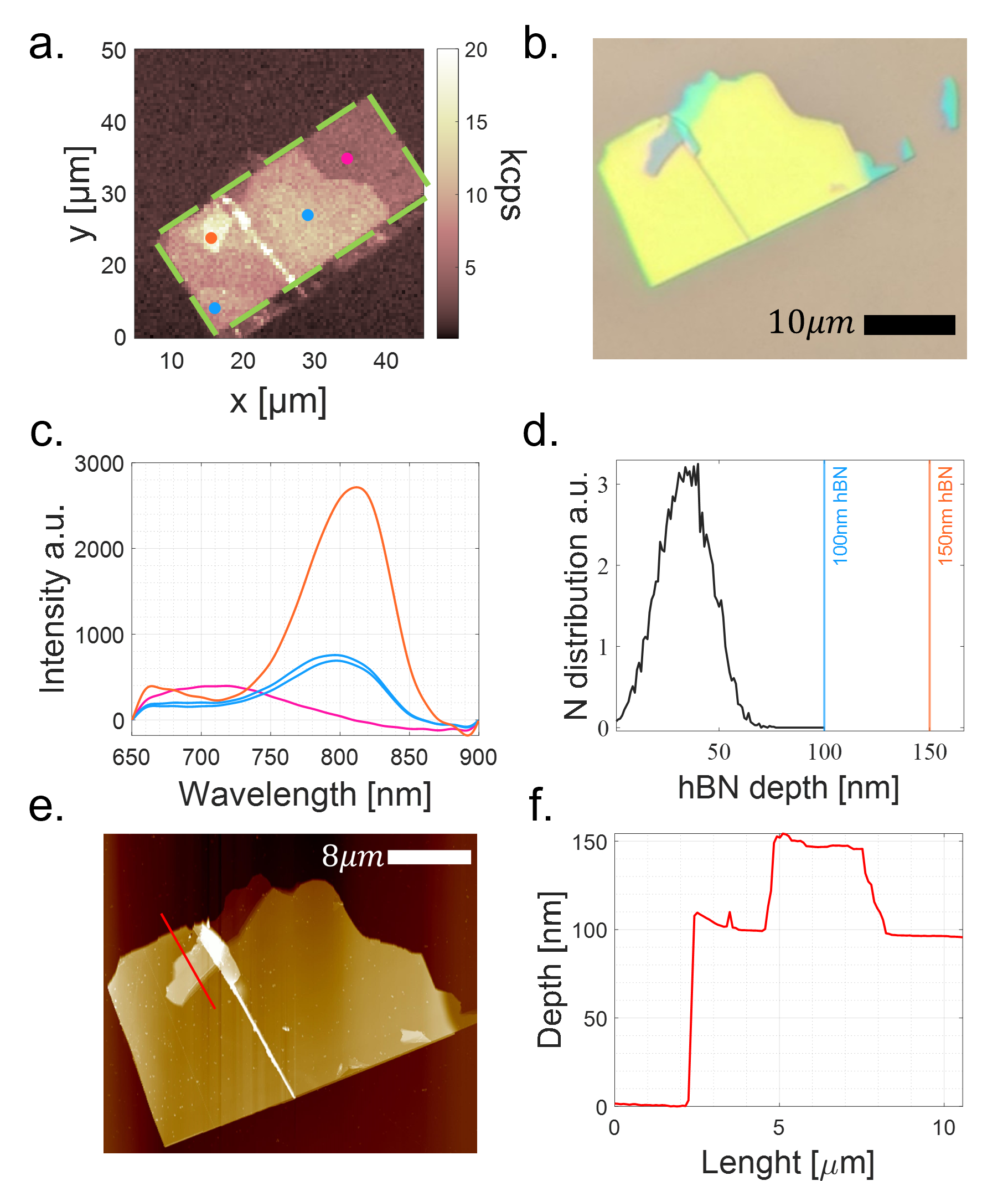}
\caption{Flake N1. (a) Confocal scan of a flake treated by N FIB implantation (Green dashed line), the dots depict the areas for which the spectral measurement was performed. (b) Microscopic image of the flake. (c) Comparison of the spectral signature of two thicknesses of the hBN - 150 nm (Orange), 100 nm (Blue) and the silicone substrate for reference (Pink). (d) SRIM simulation showing the N ion depth distribution for energy of 12 keV, $\mu=33.3 \ nm$,\ $\sigma=12.1 \ nm$. (e-f) AFM scan of the flake showing it's thickness in the relevant areas.}
\label{fig:flake_N1}
\end{figure}

A uniform fluence of Nitrogen FIB treatment was applied on flake N1 [as seen in Figure \ref{fig:flake_N1}(a)], which consists of regions with varying thicknesses [100-150 nm as measured by atomic force microscope (AFM), see Figures \ref{fig:flake_N1}(e-f)]. FIB treatment was carried out using the following parameters:

\begin{itemize}
\item beam size 0.01${\mu}m^2$,
\item dwell time 1${\mu}s$, 
\item current of 10$pA$,  
\item exposure repetitions 1k.
\end{itemize}

These parameters translate to a dose of $6.25{\cdot}10^{14}$ $cm^{-2}$.
Spectral measurements confirm the creation of VB$^-$s in flake N1 [Figure \ref{fig:flake_N1}(c)].

The spectral measurements depicted in the figures were smoothed out using triangle filters for visual purposes. Additionally, the spectral raw data were fitted to gaussians in the relevant areas, to estimate the SNR from the VB$^-$s, further details of the SNR calculation can be found in the supplementary materials \cite{suppRef}).
The strongest PL signal and best SNR were observed from the thicker area of the flake (150 nm, SNR $\sim$20), compared to a thinner area (100 nm, SNR $\sim$9) [see Figure \ref{fig:flake_N1}(c)]. 
While some area of the Si substrate was exposed as well, and exhibited some photo-luminescence (PL), it was clearly unrelated to the relevant hBN defects, as can be observed from the spectral signature [Figure \ref{fig:flake_N1}(c)].

\begin{figure} [tbh!]
    {\includegraphics[width = 1 \linewidth, trim={0 1cm 0 0}]{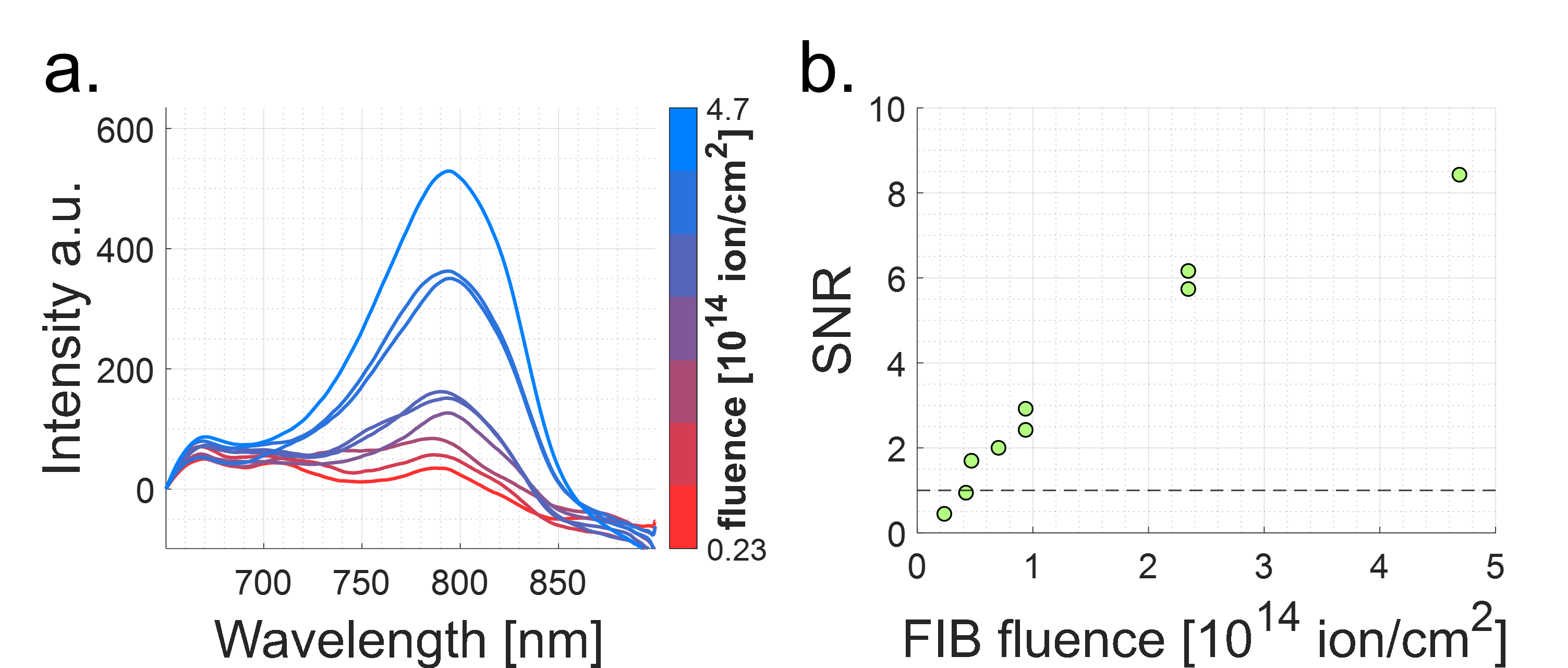}}
    \caption{Flake O2. (a) Oxygen FIB dose test(see colorbar) spectral measurement results.
    (b) SNR of the VBs spectral signal for different dosages, for a given thickness of 83 nm and implantation energy (12 keV), the SNR grows with the dose, starting to approach saturation, though not demonstrated.}
\label{fig:oxygen_dose_test}
\end{figure}

Figure \ref{fig:oxygen_dose_test}(a) depicts dose test experiments performed with Oxygen on flake O2 which is 83 nm thick, with the following parameters: energy 12 keV, beam size 0.04 ${\mu}m^2$, dwell times of 1 ${\mu}s$ and 10 ${\mu}s$, current of 30 $pA$ and varying repetitions.
The PL intensity increases with dose, while different dwell times don't have a significant impact on PL intensity for the same dose.

\subsection{Thickness Dependence}

\begin{figure}[tbh]
    {\includegraphics[width = 1 \linewidth, trim={0 0.25cm 0 0}]{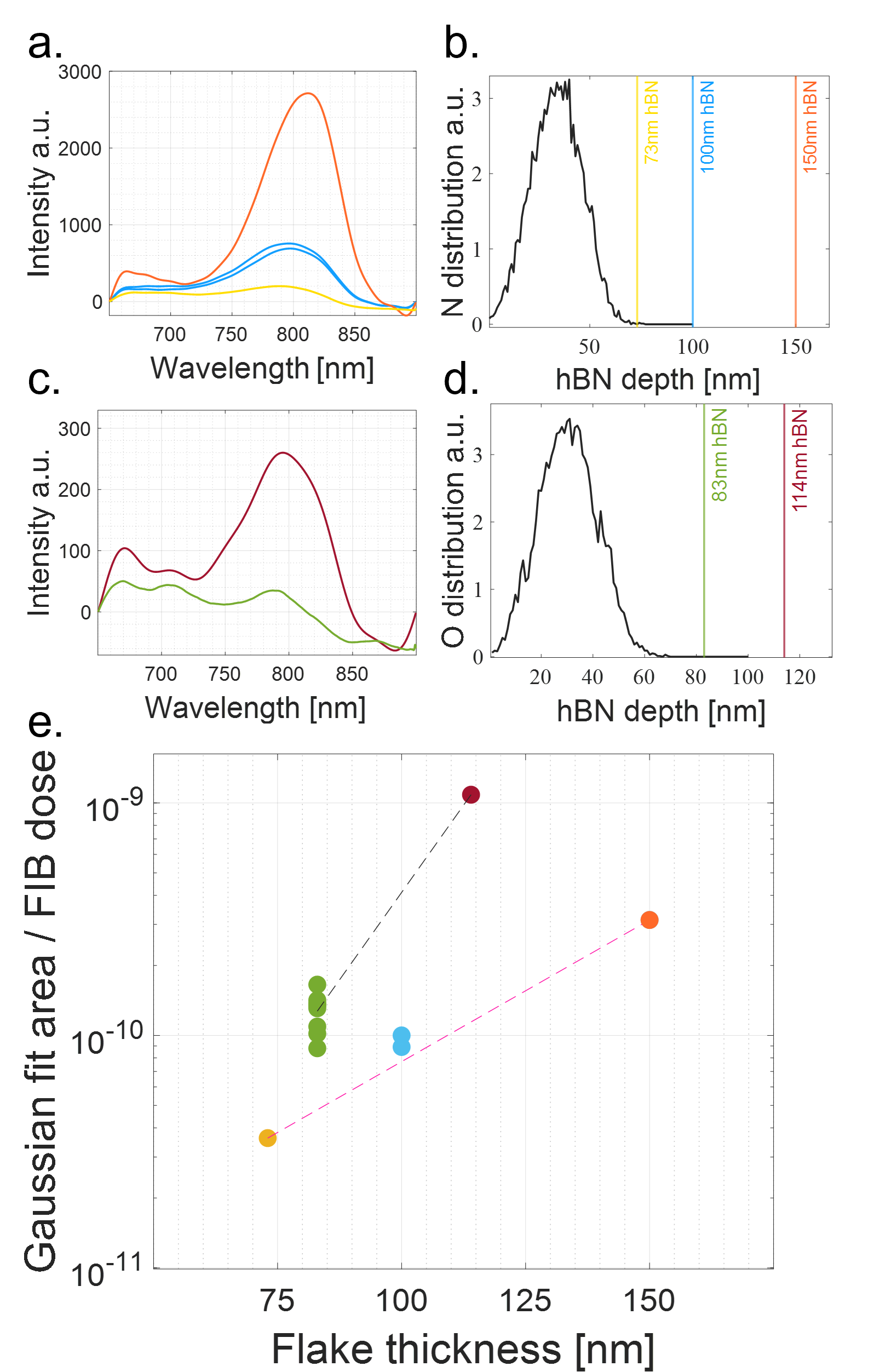}}
     \caption{Flake N1 (Blue and Orange) and flake N2 (Yellow), flake O1 (Red) and flake O2 (Green). (a) spectral measurement comparison for similar doses of Nitrogen FIB ${\sim}5{\cdot}10^{14}cm^{-2}$, the thicker the flake the higher the PL intensity. (b) SRIM simulation for Nitrogen at 12 keV alongside the different flake thicknesses.
     (c) Spectral measurement of hBN flakes after Oxygen FIB treatment with similar dose ${\sim}2{\cdot}10^{13}cm^{-2}$, the thicker flake, O1, has significantly stronger PL intensity than the thinner flake, O2. 
     (d) SRIM depth range simulation for O ions on hBN, at energy of 12 keV, $\mu=29.9 \ nm$,\ $\sigma=11.1 \ nm$. (e) thickness dependence on the spectral signal of the VB$^{-}$s (divided by the ion fluence), where the thicker flakes show stronger signal for both ion species Nitrogen (dashed Pink) and Oxygen (dashed Black).}
\label{fig:thickness_ON}
\end{figure}

SRIM simulations [Figures \ref{fig:thickness_ON}(b), \ref{fig:thickness_ON}(d)] predict the maximal ion penetration depth to be approximately 70 nm (for our parameters). Therefore it is not a-priori clear why for a given ion beam fluence, using 12 keV, a flake thicker than 70 nm, would show stronger PL and better SNR.
Therefore we conjecture that this result suggests that mainly ion channeling or secondary ions create the VB$^-$ defects, rather than first collision ions. Similar behavior have been observed for NV$^-$s in diamonds \cite{NV_SRIM_doi:10.1021/nl102066q}, which highlights the limitation of SRIM to predict the depth of defect created in the material.

For the Nitrogen FIB we see a clear difference in PL intensity [Figure \ref{fig:thickness_ON}(a)] and in calculated SNR, as a function of flake thickness [Figure \ref{fig:thickness_ON}(b)].
Similarly, Oxygen FIB implantation spectral measurements [Figure \ref{fig:thickness_ON}(c)] indicate that the thicker flake O1 has greater SNR of about 4 while for the thinner flake, O2, the SNR was less than 0.5, under similar FIB fluence. 

Figure \ref{fig:thickness_ON} shows that for the thinner (83 nm) flake, O2, a PL intensity of $\sim 250$ in the arbitrary, but fixed, intensity units, requires a higher dosage to match the same PL that was measured for the thicker (114 nm) flake, O1. From Figure \ref{fig:oxygen_dose_test}(b) it can be seen that an increase of one order of magnitude in the fluence for the thinner flake is needed to match the same SNR of the thicker flake.

These results confirm the argument that a simple SRIM range simulation does not predict the depth of VB$^-$s in hBN, and potentially indicates the non-negligible role of secondary ions in creating these defects. Practically, to achieve a specific VB$^-$ concentration in a given hBN flake, for a certain beam energy, the implantation dose should be calibrated with respect to the flake's thickness, since thicker flakes require a lower ion dose, compared to thinner ones.

\subsection{Optically detected magnetic resonance}

The VB$^-$, like the NV$^-$, is a spin one system with a triplet ground state. Its resulting zero field splitting is $\sim$3.47 GHz \cite{vbOdmr, Ivady2020}. As such, and given its optical response, an Optically Detected Magnetic Resonance (ODMR) measurement can be performed to observe and characterize the VB$^-$ spin properties \cite{interestingVBodmr}.

The pulsed ODMR sequence consists of an optical pumping pulse to initialize the spins to their $m_s = 0$ state, a microwave (MW) pulse applied with varying frequencies, followed by a final optical pulse for spin readout. At the spin resonant MW frequency spin transitions are induced between the $m_s = 0$ and $m_s = \pm 1$ ground states, such that the resulting fluorescence is suppressed. 

\begin{figure}[tbh]
{\includegraphics[width = 0.9 \linewidth, trim={0 0.25cm 0 0}]{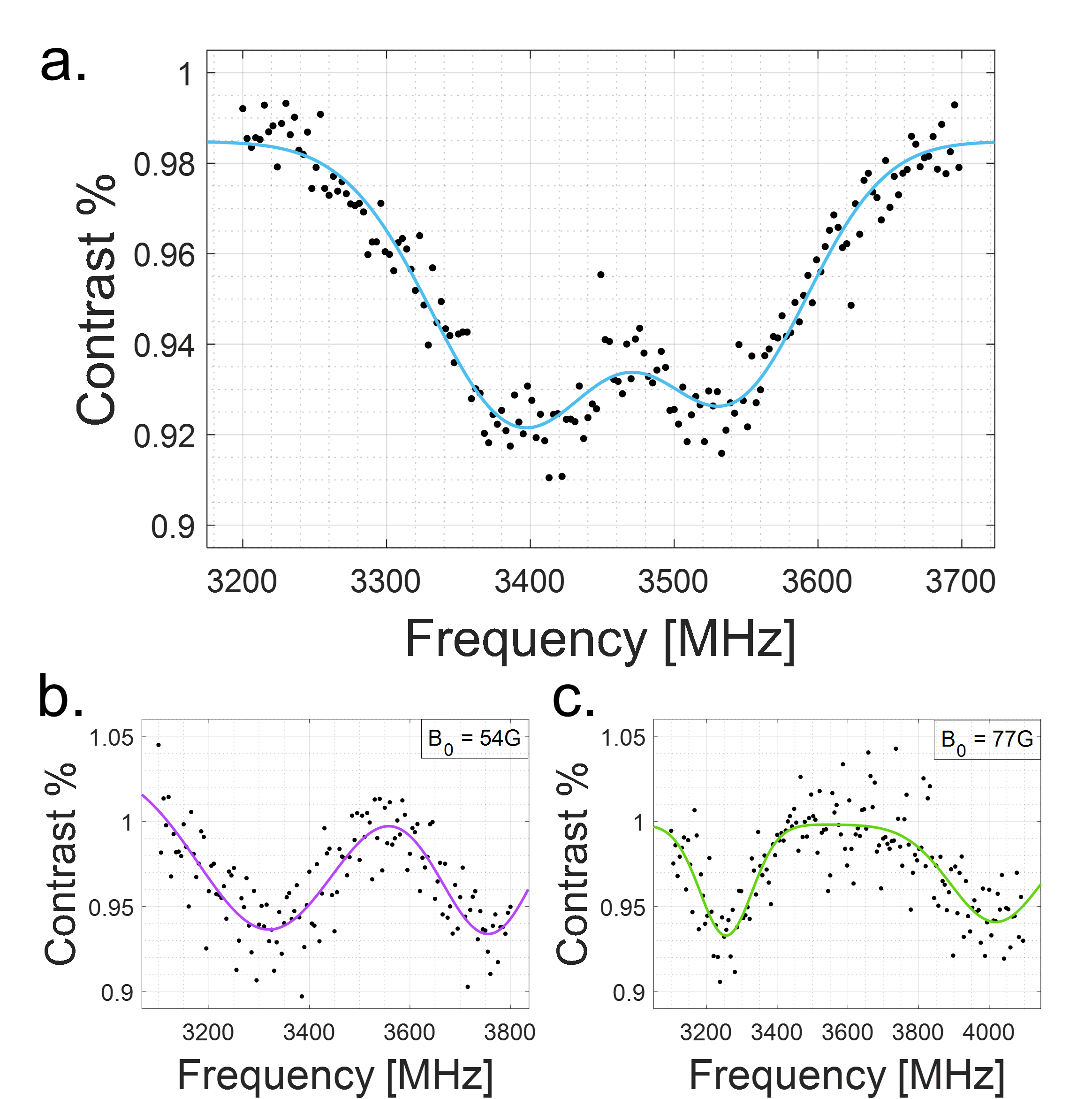}}
\caption{Measured results for flake N1, ODMR with a bias field, showing the Zeeman shift of $m_s =\pm 1$, this is consistent with the expected $\gamma=2.8 \ \frac{MHz}{G}$ of a free electron. (a) no Bias field. (b) Bias field of 54 G and (c) 77 G.}
\label{fig:ODMR_DC_magnetic_field}
\end{figure}

To perform the ODMR we used an $\Omega$-shaped co-planar antenna, ensuring that the flake is as close as possible to the conducting loop to achieve higher effective MW power (and thus stronger driving). 
Zero field ODMR measurements on the nitrogen FIB implanted flake are presented in figure \ref{fig:ODMR_DC_magnetic_field}(a). 
ODMR with an external magnetic field at 54 Gauss and 77 Gauss are presented in figure \ref{fig:ODMR_DC_magnetic_field}(b,c), demonstrating the linear Zeeman split between the $m_s = \pm1$ ground spin states.

Similar results were achieved on Oxygen FIB flakes (see supplementary material), exhibiting ODMR contrast variations with PL intensity on the different flakes, related to different VB$^-$ concentrations.

\section{Summary}
In summary, we have presented a study of the application of Nitrogen and Oxygen FIB implantation for creating VB$^-$ defects in hBN flakes. The field of atomic defects in 2D materials is growing rapidly, with clear prospects for potential impact on various quantum technologies, including sensing and photonics. Nevertheless, a systematic study of the processes and techniques for the robust creation of desired defects is missing, and is the focus of this manuscript. Our results show that both Oxygen and Nitrogen ions at 12 keV can efficiently create VB$^-$s in hBN flakes. We find that hBN flake thickness plays a key role in the success of the process and in the achievable VB$^-$ density, and for a given beam energy, a higher dose is required for thinner flakes. Furthermore, we find that the cleaning process used on the flakes prior to implantation is very important. Specifically, the inclusion of a forming gas heating step or a simple hot plate cleaning proved crucial for successful creation of VB$^-$ in hBN.

We also note that in certain cases flakes could be created by well-controlled, clean methods not related to exfoliation \cite{hbnGrowth1, hbnGrowth2}. Nevertheless, exfoliation from bulk crystal is a widely used technique, with the advantage of providing access to high quality material. As such, the techniques presented here, and specifically the cleaning processes, should be carefully considered prior to implantation.

\section{Acknowledgements}

N.B. acknowledges support from the European Union’s Horizon 2020 research and innovation program under Grant Agreements No. 101070546 (MUQUABIS) and No. 828946 (PATHOS), and has been supported in part by CIFAR, the Ministry of Science and Technology, Israel, the innovation authority (Project No. 70033), and the ISF (Grants No. 1380/21 and No. 3597/21).
G.H. acknowledges support by the Melbourne research scholarship.
I.A. acknowledges financial support from the Australian Research Council (CE200100010, FT220100053) and the Office of Naval Research Global (N62909-22-1-2028).
We acknowledge financial support from the Danish Agency for Higher Education and Science, grant no. 9096-00023B.

\bibliography{VB_FIB_REF.bib}

\section{APPENDIX A - hot plate}
As mentioned in the manuscript, some samples were cleaned using simple hot plate, set to $400^{\circ}C$ for 15 minutes.
\begin{figure}[H]
    \centering
    \includegraphics[width = 0.9 \linewidth]{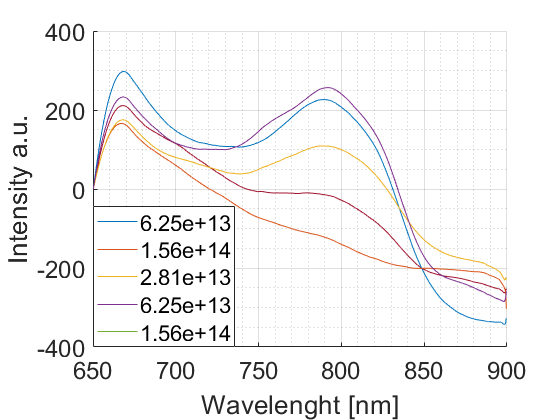}
    \caption{Flake N3 dose test, varying fluences in the range of $10^{13}\sim10^{14} cm^{-2}$}
    \label{fig:hot_plate}
\end{figure}
In Figure \ref{fig:hot_plate} the spectral signature of the FIB exposed areas, suggest that VB$^-$s were created in flake N3. The hot plate cleaning process did not eliminate all contaminants from the flakes, thus some fluorescence is visible for wavelengths below $700 \ nm$, similarly to flake that went though annealing in forming gas.
We propose that the contaminants might have been introduced during a SEM imaging session, though not thoroughly tested.
\section{APPENDIX B - data smoothing and SNR}
\begin{figure}[H]
    \centering
    \includegraphics[width = 0.9 \linewidth]{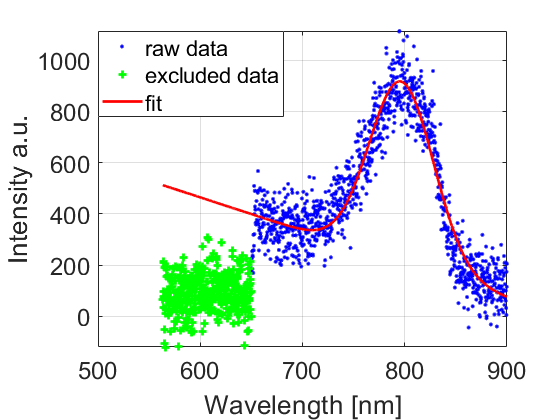}
    \caption{Spectral measurement from flake N1 with a fit to the following model: $A\cdot\exp{(-\frac{(\lambda-\mu)^2}{2\cdot\sigma^2})}+m\cdot{x}+C$}
    \label{fig:gaussian_fit}
\end{figure}

For clear visibility of the spectral measurements, the raw data was smoothed out using a triangular filter as described in the following manner:
\begin{equation}%
\large 
S(\lambda) = {\frac{\displaystyle\sum_{-\lfloor{\frac{w}{2}}\rfloor}^{\lfloor{\frac{w}{2}}\rfloor}({\lceil{\frac{w}{2}}\rceil}-{\lvert{m}\rvert})Y(\lambda+m\cdot\Delta\lambda)}{w+2\cdot\sum(w-j)}}
\end{equation}
Where $S(\lambda)$ is the smoothed out signal, $\Delta\lambda$ the wavelength increments, $Y(\lambda)$ the raw signal and $w$ is the smoothing size of the filter.
Then to estimate an SNR value, a gaussian shape was extracted from the following fit on the raw data of the spectral measurement (as can be seen in Figure \ref{fig:gaussian_fit}) using the following formula:
\begin{equation}%
\large 
fit = {A\cdot\exp{(-\frac{(\lambda-\mu)^2}{2\cdot\sigma^2})}+m\cdot{x}+C}
\end{equation}
While excluding data point below $650 \ nm$, since a low pass filter of that wavelength was used during acquisition.
Finally for the SNR calculation, the ration between the area under the gaussian and root mean square error ($\sigma_{RMSE}$) of the fit,
as shown here:
\begin{equation}%
\large 
SNR = {\frac{A\cdot\int{\exp{(-\frac{(\lambda-\mu)^2}{2\cdot\sigma^2})}}d\lambda}{\sigma_{RMSE}^2}}
\end{equation}
\section{APPENDIX C - electron irradiation}
\begin{figure}[H]
    \centering
    \includegraphics[width = 0.99 \linewidth]{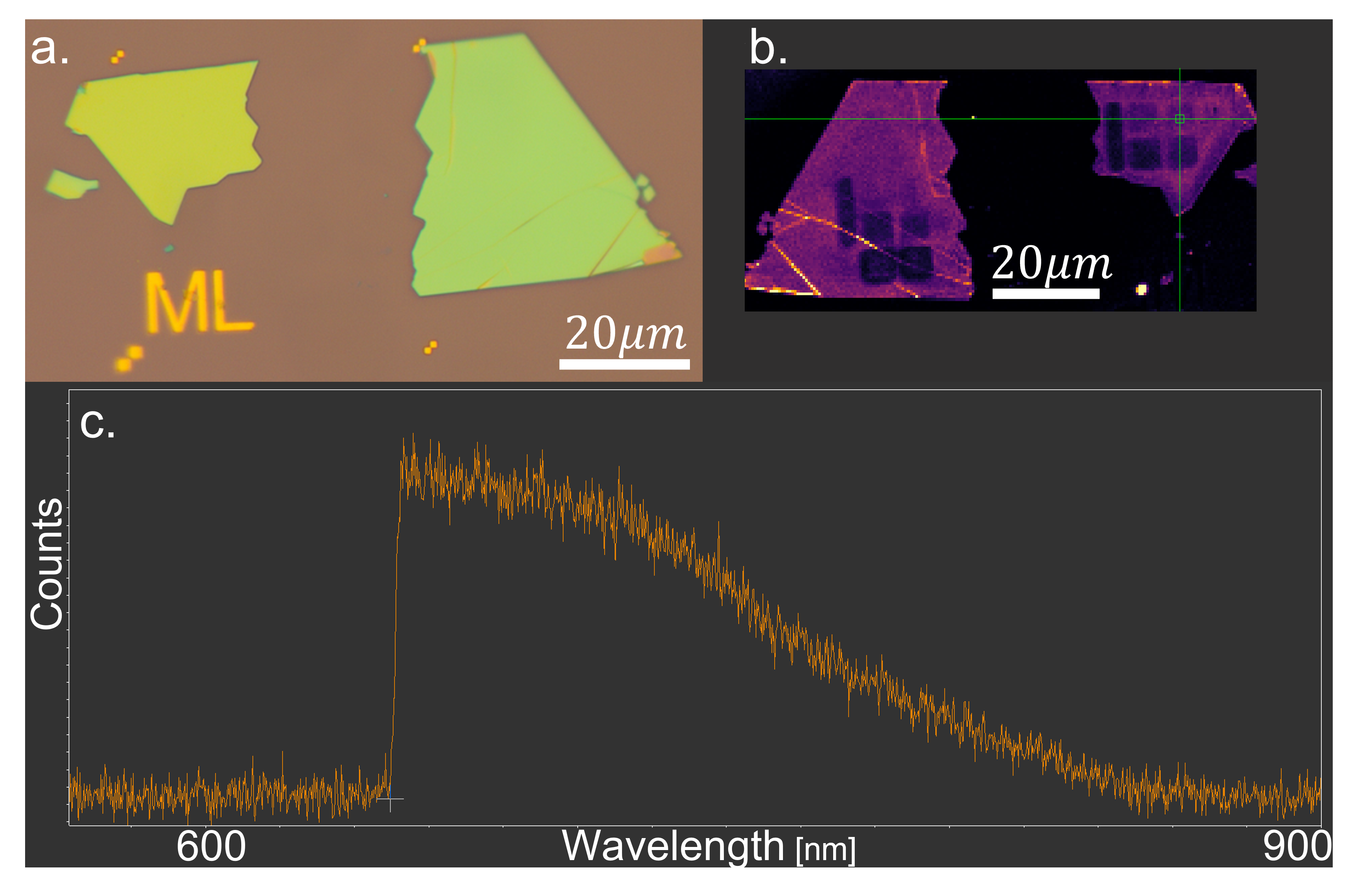}
    \caption{Two flakes that underwent e-beam dose test ranging from $10^{17}cm^{-2}$ to $10^{19}cm^{-2}$ at 100 keV. (a) microscope image of the flakes (b) confocal scan of the fluorescence from the flakes after the irradiation. (c) spectral wavelength of exposed areas on the flakes. }
    \label{fig:eBeam}
\end{figure}

\begin{figure}[H]
    \centering
    \includegraphics[width = 0.9\linewidth]{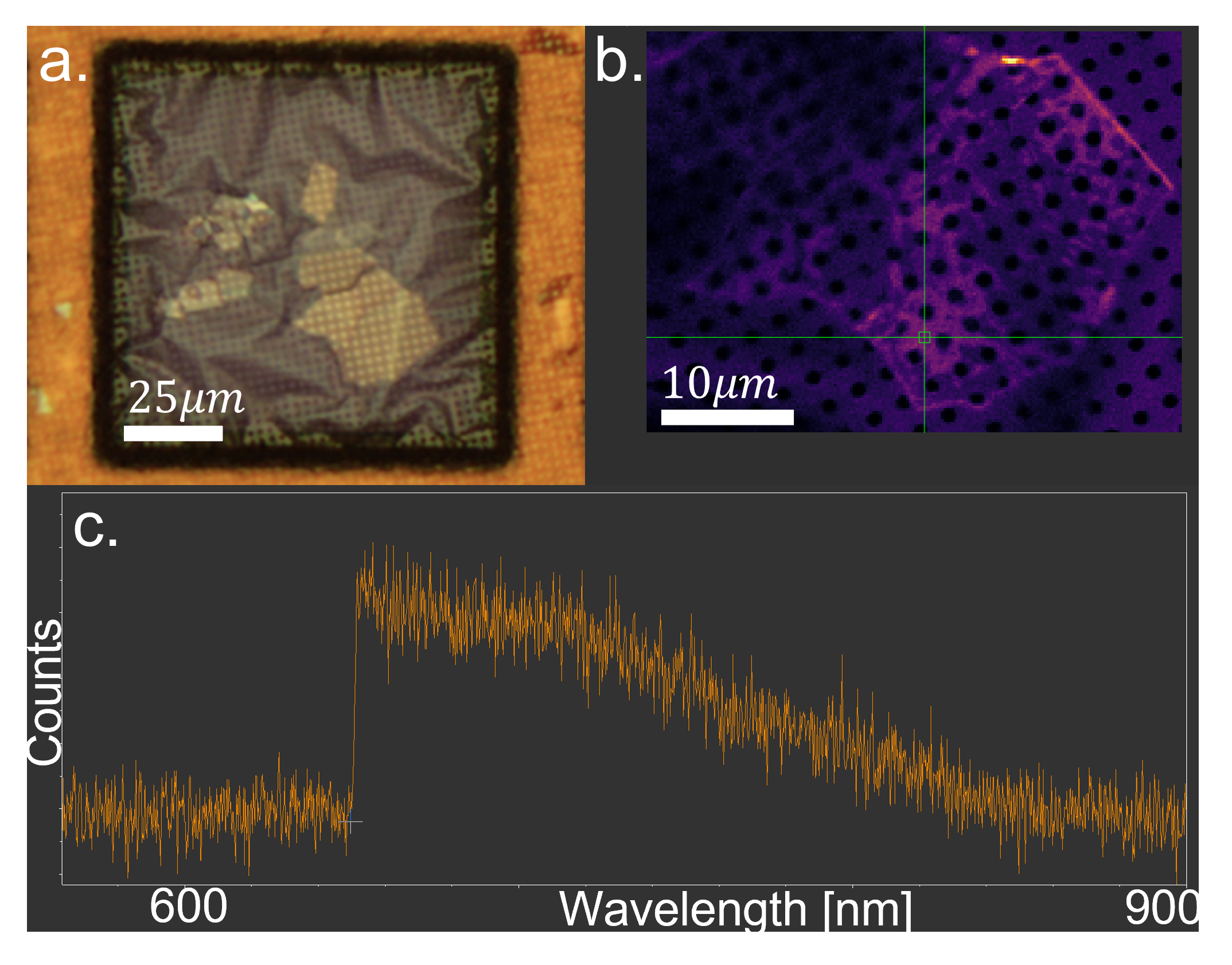}
    \caption{Two flakes that underwent TEM dose test ranging from $10^{19}cm^{-2}$ to $10^{22}cm^{-2}$ at 300 keV. (a) microscope image of the flakes (b) confocal scan of the fluorescence from the flakes after the irradiation. (c) spectral wavelength of exposed areas on the flakes.}
    \label{fig:TEM}
\end{figure}

There were few attempts to create VB$^-$s using electron irradiation with different energies, using two different techniques, e-beam (at 100 keV) and TEM (at 300 keV), the examples shown in Figures \ref{fig:eBeam} and \ref{fig:TEM}, these attempts weren't successful as can be seen in the spectral measurements(Figure \ref{fig:eBeam}.c and \ref{fig:TEM}.c) regardless of annealing treatment. 
\section{APPENDIX D - ion implantation}
\begin{figure}[H]
    \centering
    \includegraphics[width = 0.99 \linewidth]{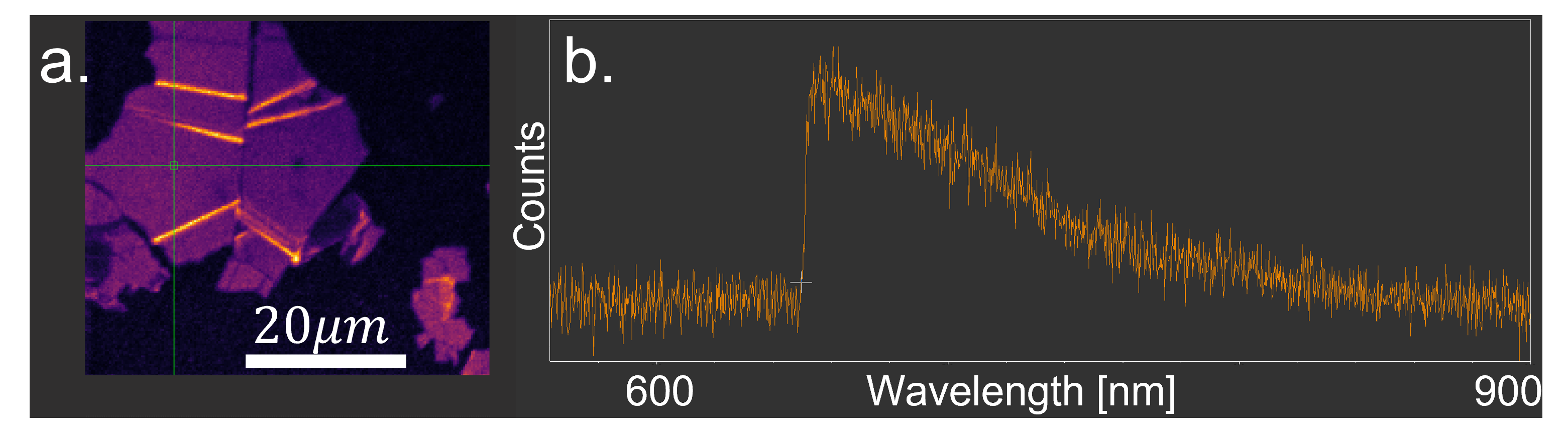}
    \caption{Two flakes that underwent uniform fluence of ion beam $10^{14}cm^{-2}$ at 40 keV. (a) confocal scan. (b) spectral measurement on flake}
    \label{fig:innovion}
\end{figure}

There was an attempt to create VB$^-$s by ion implantation using commercial (semiconductor industry) ion implanter instrumentation (Innovion Corp.). This was unsuccessful as can be seen from Figure \ref{fig:innovion}.b, but further note that in this case the flakes were not properly cleaned prior to ion treatment.
\section{APPENDIX E - ODMR}
\begin{figure}[H]
    \centering
    \includegraphics[width = 0.89 \linewidth]{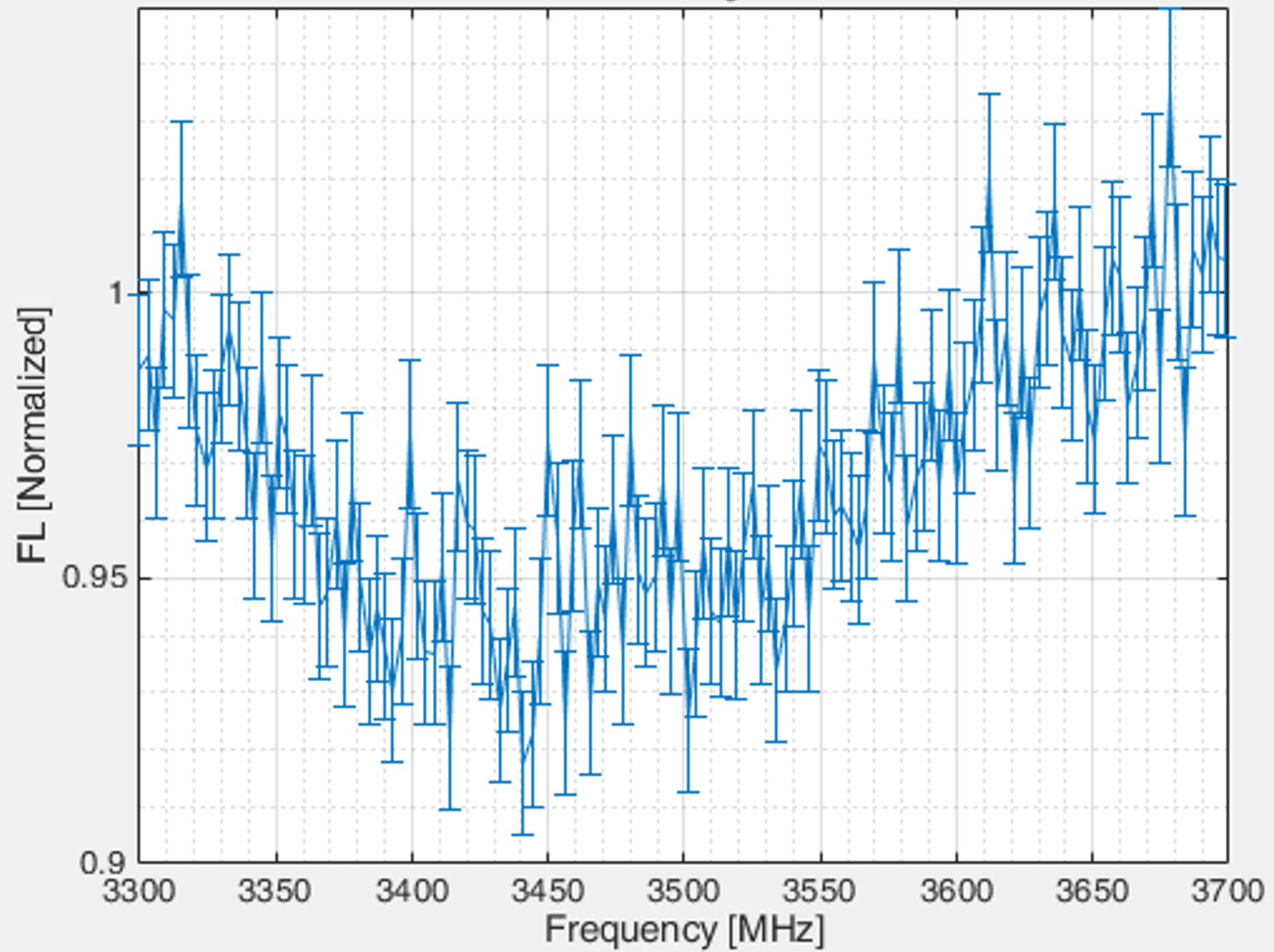}
    \caption{Flake O1, ODMR measurement without bias magnetic field}
    \label{fig:Oxygen_ODMR}
\end{figure}


\end{document}